# MULTIPACTING PROCESSING IN CRYOMODULES FOR LCLS-II AND LCLS-II-HE[*]

A. Cravatta[†], T. Arkan, D. Bafia, J. Kaluzny, S. Posen
Fermi National Accelerator Laboratory, Batavia, IL, USA
J. Vennekate, M. Drury, Thomas Jefferson National Accelerator Facility, Newport News, VA, USA
S. Aderhold, M. Checchin, D. Gonnella, J. Hogan, J. Maniscalco, J. Nelson, R. Porter, L. Zacarias
SLAC National Accelerator Laboratory, Menlo Park, CA, USA

## Abstract

Multipacting (MP) is a phenomenon which can affect stability in particle accelerators and limit performance in superconducting radio frequency cavities. In the TESLA shaped, 1.3 GHz, 9-cell cavities used in the LCLS-II (L2) and LCLS-II-HE (HE) projects, the MP-band (~17-24 MV/m) lies within the required accelerating gradients. For HE, the operating gradient of 20.8 MV/m lies well within the MP-band and cryomodule testing has confirmed that this is an issue. As such, MP processing for the HE cryomodule test program will be discussed. Early results on MP processing in cryomodules installed in the L2 linac will also be presented, demonstrating that the methods used in cryomodule acceptance testing are also successful at conditioning MP in the accelerator and that this processing is preserved in the mid-term.

## MULTIPACTING IN LCSL-II AND LCLS-II-HE CAVITIES

The phenomenon of multipacting (MP) in SRF cavities has been described in many places and one such treatment can be found in Ref. [1]

In the TESLA shaped, 1.3 GHz, 9-cell cavities of the type used in LCSL-II (L2) and LCSL-II-HE (HE) cryomodules (CM), the MP-band is in the range of 17-24 MV/m. This poses a particular challenge to the two projects as cavities will operate at gradients within the MP-band. For HE in particular, the nominal gradient of 20.8 MV/m lies well within this range.

Fermilab has built, tested, and delivered 22 CM for the L2 project and 14 CM are slated for HE. As of this conference, 5 CM, including a prototype, the vCM, have been successfully tested. Details of the vCM testing and the Fermilab CM testing scheme can be found in Ref. [2, 3].

## MP IN THE FERMILAB TEST STAND

Of the 40 cavities within the 5 HE CM tested at the Fermilab Cryomodule Test Facility so far, 35 have exhibited MP. It should be noted that of the 5 cavities which did not exhibit MP, 1 of these was limited to well below the MP-band due to field emission (FE). Table 1 shows any occurrence of MP in the 5 CM tested so far: the vCM, the first article CM (F21), and the 3 production CM tested to so far (F22-F24).

Table 1: Occurrence of MP During HE CM Testing

| Cavity # | vCM | F21 | F22 | F23 | F24 |
|---|---|---|---|---|---|
| 1 | YES | YES | YES | YES | YES |
| 2 | YES | NO | YES | YES | YES |
| 3 | YES | YES | YES | YES | YES |
| 4 | YES | NO | NO | YES | YES |
| 5 | YES | YES | YES | YES | YES |
| 6 | YES | YES | NO | YES | NO* |
| 7 | YES | YES | YES | YES | YES |
| 8 | YES | YES | YES | YES | YES |

A common signature of MP in the test stand is sporadic quenching in the MP-band and associated radiation spikes as seen in Fig. 3. A quench interlock in the EPICS controls system inhibits RF and produces a fault waveform, with traces of the forward (red), reverse (orange), and transmitted powers (blue) along with a trace of the calculated decay of the transmitted power signal (light blue) to verify a 'true' quench (Fig. 1) and not a false signal (Fig. 2).

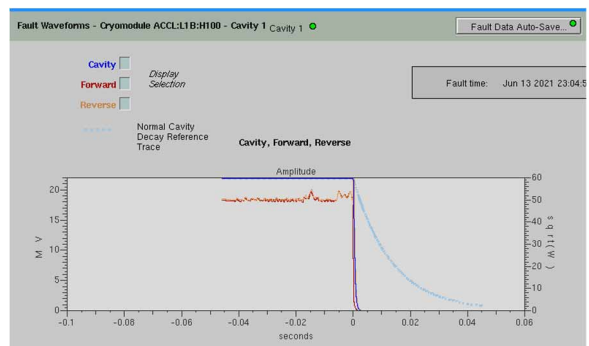

Figure 1: Waveforms showing a 'true' quench.

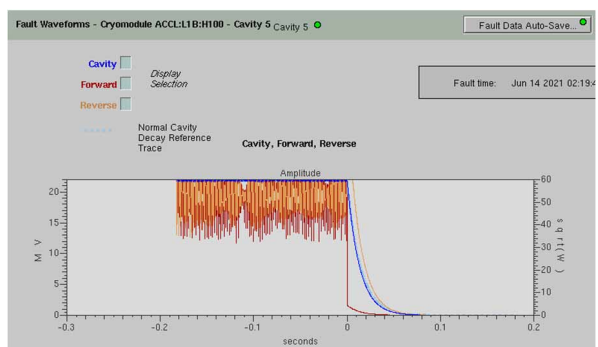

Figure 2: Waveforms showing a 'false' quench.

---

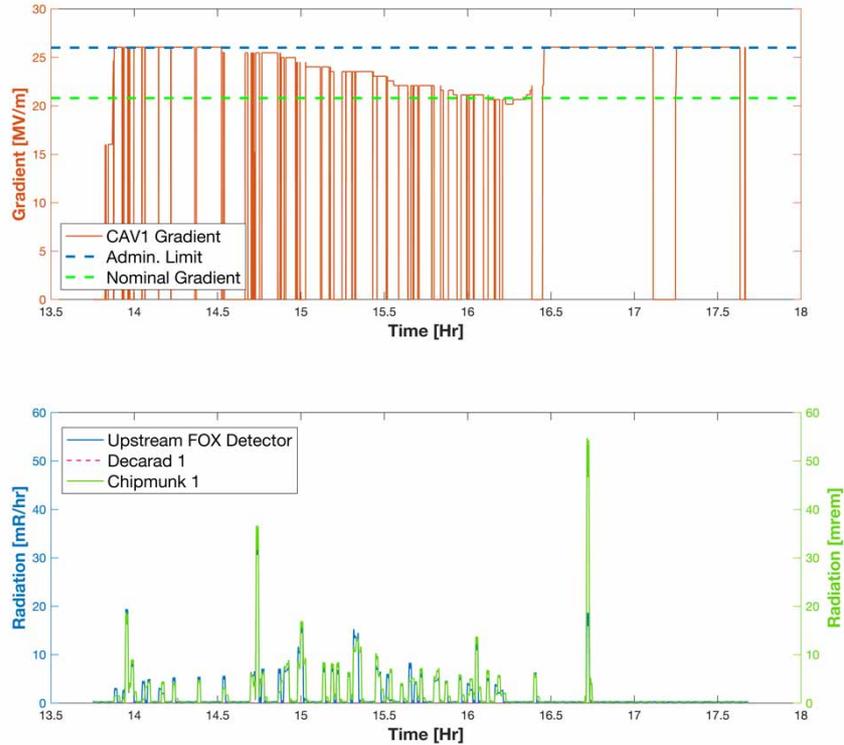

Figure 3: MP in a HE cavity. Quenching can be seen in the top plot, in which the gradient quickly falls to zero. In the bottom plot, the associated radiation can be seen on three radiation monitors.

## MP PROCESSING

MP processing consists of repeatedly quenching a cavity with minimal time (seconds) between quenches. For stubborn cavities with severe MP, pulsed processing may be required, however in a majority of cases processing is done in CW.

During processing, power is returned to the cavity after a quench within seconds. As a cavity processes, the duration between quenches increases and stability at a given gradient increases. At Fermilab, processing is done in three stages, which will be described in the next sections.

### Short-Term Processing

The first tests performed during a CM test at Fermilab include power rises to 16 and 21 MV/m as these are critical milestones within the Acceptance Criteria. Discussion of the CM test plan and Acceptance Criteria can be found in Ref. [3].

During the power rise is typically when MP is first observed. If MP is seen in a cavity prior to reaching 21 MV/m, then the cavity will be processed up to this field. Otherwise, the field in the cavity will be increased up to the admin. limit of the test stand, 26 MV/m, to record the maximum gradient of the cavity.

### Mid-Term Processing

The next stage of testing involves recording the Usable Gradient, in which the cavity must run stably, without quenching, and with field emission (FE) levels, if present, of less than 50 mR/hr, for 1 hr uninterrupted.

For cavities that did not quench or multipact during the power rise to max. gradient, and which did not reach the admin. limit in gradient, this is typically the stage in which MP processing is needed. It is not uncommon at this stage for MP to occur at the max. gradient and then be present through the rest of the MP-band, where previously the cavity had not experienced MP during the short duration at various fields during the power rise. Often a cavity at this stage will not only need MP-processing at the quench field limit, but also processing throughout the rest of the MP-band.

In increments as small as 0.1 MV, cavities can exhibit MP and quenching, with radiation spikes detected on various radiation monitors. Processing at any given field is typically on the order of tens of seconds or minutes and full processing can take 30 min to 1 hr. Some stubborn cavities may take several hours.

An example of a cavity exhibiting MP and being processed through the MP-band is shown in Fig. 4. Here, the cavity quenches around 21 MV/m and then again continuously as the field is lowered to ~19 MV/m. The gradient is then lowered until stable and finally increased in 0.1 MV increments, processing along the way. In this case, the cavity is process for ~10min up to 23 MV/m before stability is achieved. The field then reaches 26 MV/m, the admin limit, and runs stable for 1 hr to collect the Usable Gradient.

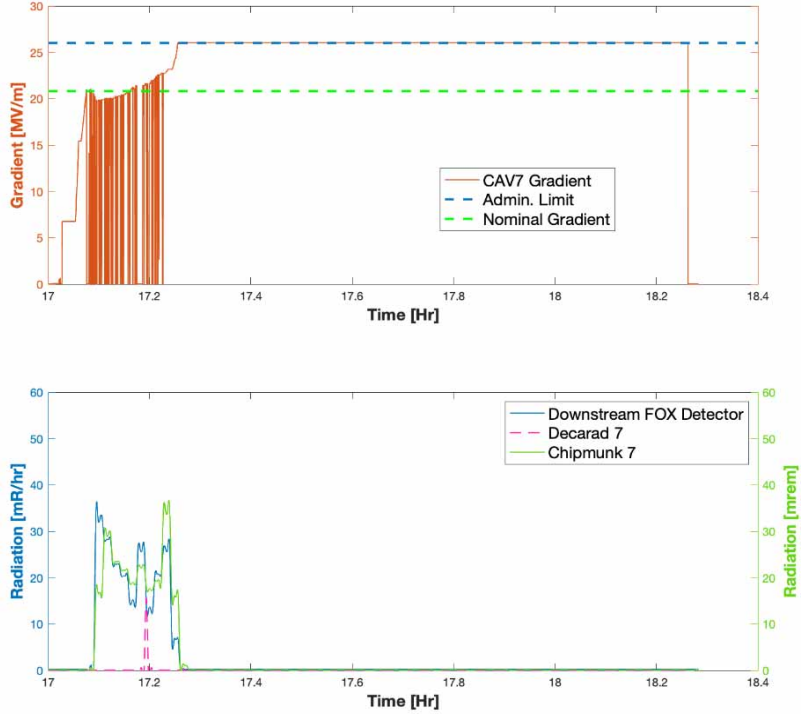

Figure 4: MP in a HE cavity is observed during the power rise. Here the MP is processed in ~10min, after 58 quenches and then a 1 hr Usable Gradient run is collected.

### Long-Term Processing

$Q_0$ measurements are taken after the bulk of MP processing is completed and both Max and Usable Gradients are collected. To recover a high $Q_0$ ($> 2.7 \times 10^{10}$), the CM undergoes a thermal cycle and fast cooldown, in which the CM is warmed to ~50 K and then cooled at a rate of >32 g/s in He mass flow after all processing is complete in an attempt to expel trapped flux, which can be a result of the quenches induced by MP [4]. A detailed study of thermal cycle/fast cooldown, $Q_0$ measurements, and $Q_0$ degradation from the prototype HE CM, the vCM, can be found in Ref. [2].

It has been observed that while some cavities may reach the admin. limit for gradient during the power rise without quenching, as well as running for 1 hr without MP/quenching at this limit, once the gradient is lowered to gradients within the MP-band, MP and quenching can often occur. $Q_0$ measurements are collected at the nominal gradient, 20.8 MV/m, and must remain stable for up to 1 hr. It is therefore prudent to process at this field prior to the thermal cycle/fast cooldown. Additionally, a Unit Test is performed as the final stage of testing, in which all cavities are run simultaneously while maintaining an average gradient at the nominal gradient for 10 hr continuously.

For this reason, long-term MP processing is performed in which cavities are run at nominal gradient for 2-4 hr, often concurrently with 3-4 cavities. This final processing has yielded positive results as seen in Table 2. Of the 35 cavities tested so far which have exhibited MP, only 4 have shown MP after processing and during the $Q_0$ measurements.

Cavities 2 and 8 from CM F23 had MP-induced quenches during the $Q_0$ measurements, however this CM required a warmup to room temperature for a minor repair and, due to time constraints in the testing schedule, the cavities could not be re-processed through the full MP-processing cycle. 3 of the 5 CM tested have been warmed to room temp. and in all cases, it has been observed that MP was reintroduced to some extent.

Table 2: Occurrence of MP During $Q_0$ Measurements and after Processing

| Cavity # | vCM | F21 | F22 | F23 | F24 |
|---|---|---|---|---|---|
| 1 | NO | NO | NO | NO | NO |
| 2 | NO | NO | NO | YES* | NO |
| 3 | NO | NO | NO | NO | NO |
| 4 | YES | NO | NO | NO | NO |
| 5 | NO | NO | NO | NO | NO |
| 6 | NO | NO | NO | NO | NO |
| 7 | NO | NO | NO | NO | YES |
| 8 | NO | NO | NO | YES* | NO |

Excluding the vCM, which underwent an atypical testing scheme with respect to production CMs, including two warmups to room temp., the first article (F21) and production CMs (F22-F24) had just one cavity quench due to MP during the 10 hr Unit Test. This was in F23, a CM that was not fully reprocessed after a warmup to room temp.

## LCLS-II COMMISSIONING

Commissioning of the 37 CM installed in the LCLS-II linac began in 2022. The nominal gradient, 16 MV/m, and admin. limit for most of the cavities, 18 MV/m, lie within the lower edge of the MP-band and MP was indeed observed.

The processing techniques developed and tested by the HE team were applied in the linac and of the 37 cavities that were processed, 100% of the cavities were able to achieve some level of processing with an average gradient gain of ~3 MV/m. Additionally, over the course of 3 months, the processing was preserved in all cavities. One such example of MP processing in the linac can be seen in Fig. 5.

Further MP processing is planned in the future as well as studies on the persistence of MP in the linac and the preservation of processing. Likewise, these techniques will be applied to the HE CMs once installed in the linac.

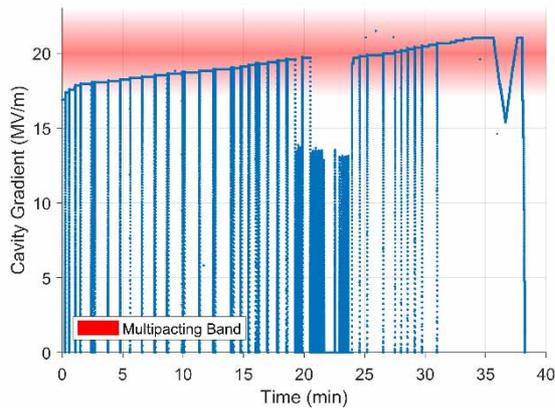

Figure 5: MP processing in the LCLS-II linac.

## CONCLUSION

Multipacting processing procedures have been developed by the LCLS-II-HE team and tested on HE cryomodules at the Fermilab Cryomodule Test Facility. Due to the operating gradients of the L2 and HE cavities lying within the MP-band, it is critical to process cavities in order to maximize stability and performance and minimize downtime.

MP has been observed in 35 of 40 HE cavities in the 5 CM tested for the HE project at Fermilab. Of the 40 processed cavities, 36 have been successfully processed up to or greater than the nominal gradient of 20.8 MV/m without recurring MP-induced quenches.

These techniques have been further implemented in the LCLS-II linac commissioning. An average gain in gradient of ~3 MV/m has been observed in 37 processed cavities and processing was retained over the course of a 3-month period.

## REFERENCES

[1] H. Padamsee, J. Knobloch, T. Hays, *RF Superconductivity for Accelerators*, Wiley Publishing, 1998.

[2] S. Posen *et al.*, "High gradient performance and quench behavior of a verification cryomodule for a high energy continuous wave linear accelerator", *Phys. Rev. Accel. Beams*, vol. 25, p. 042001, 2022.
 doi:10.1103/PhysRevAccelBeams.25.042001

[3] A. Cravatta *et al.*, "LCLS-II-HE Cryomodule Testing at Fermilab", in *Proc. LINAC'22*, Liverpool, UK, Aug.-Sep. 2022, pp.721-723.
 doi:10.18429/JACoW-LINAC2022-THPOJO12

[4] S. Posen *et al.*, "The role of magnetic flux expulsion to reach $Q_0 > 3 \times 10^{10}$ in superconducting rf cryomodules", *Phys. Rev. Accel. Beams*, vol. 22, p. 032001, 2022.
 doi:10.1103/PhysRevAccelBeams.22.032001